\documentclass[prd,aps,floatfix,nofootinbib,twocolumn,tightenlines,showpacs]{revtex4}
\usepackage{dcolumn}
\usepackage{amsmath}
\usepackage{latexsym}
\usepackage{graphicx}
\usepackage{bm}

\usepackage{color}

\def\svev#1{\left\langle #1\right\rangle}       
\def\tr{{\rm tr}\,}
\def\Tr{{\rm Tr}\,}

\long \def \blockcomment #1\endcomment{}
\def\det{{\rm det}}

\def\Eq#1{Eq.~(\ref{#1})}
\def\tbeta{\tilde\beta}
\def\csw{c_{\text{SW}}}
\def\Sl#1{\rlap{\hbox{$\mskip 3 mu /$}}#1}      

\def\Dslash{\Sl D}

\newcommand{\bee}{\begin{equation}}
\newcommand{\ee}{\end{equation}}
\newcommand{\beea}{\begin{eqnarray}}
\newcommand{\eea}{\end{eqnarray}}

\def\ttl#1{``#1''}
\def\tK{\tilde{K}}

\begin{document}

\title{Non-perturbative beta function in three-dimensional electrodynamics}
\author{Ohad Raviv}
\email{ohad.raviv@gmail.com}
\author{Yigal Shamir}
\author{Benjamin Svetitsky}
\affiliation{Raymond and Beverly Sackler School of Physics and Astronomy, Tel~Aviv University, 69978 Tel~Aviv, Israel}

\begin{abstract}
We apply the Schr\"odinger functional method to the Abelian gauge theory in three dimensions with $N_f=2$ four-component fermions.
We find that the calculated beta function does not cross zero in the range of coupling we study.
This implies that the theory exhibits confinement and mass generation, rather than a conformal infared regime.
\end{abstract}

\maketitle

\section{Introduction}
The infrared structure of massless quantum electrodynamics in three dimensions (QED3) is an old problem~\cite{Pisarski:1984dj}.
Like other three-dimensional theories, QED3 is plagued by
infrared divergences.
It is widely believed that these signal a ground state that is far from the perturbative vacuum.
Two possibilities have been entertained~\cite{Appelquist:2004ib}.
One is that the infrared physics of the theory is described by confinement of
charge, with a concomitant dynamical mass for the fermions.
The other is that dynamical charges screen the Coulomb interaction, so that charges at large distances interact via a power law potential.
In the first case the scale of the theory's physics is presumably set by the coupling $e^2$, which has dimensions of mass.
In the second case, there is no scale in the infrared physics, but rather an
emergent conformal symmetry.
A number of studies have concluded that confinement and mass generation occur
if the number of (four-component) fermions, $N_f$, is small.
It is clear, on the other hand, that in the large-$N_f$ domain the theory is
screened and conformal.
Estimates of $N_c$, the critical value of $N_f$ that divides the two domains,
have varied widely in various analytical approaches \cite{Appelquist:1986fd,Appelquist:1988sr,Nash:1989xx,Pennington:1990bx,Curtis:1992gm,Appelquist:1994ui,Ebihara:1994wm,Maris:1996zg,Aitchison:1997ua,Gusynin:1998kz,Appelquist:1999hr,Gusynin:2003ww,Fischer:2004nq,Kaveh:2004qa,Bashir:2008fk,Goecke:2008zh,Braun:2014wja}
and lattice simulations \cite{Dagotto:1988id,Dagotto:1989td,Azcoiti:1993ey,Alexandre:2001pa,Hands:2002dv,Lee:2002id,Hands:2004bh,Fiore:2005ps,Strouthos:2008kc}
(for interim summaries see~\cite{Appelquist:2004ib,Strouthos:2008kc}).

We can explore this question via the renormalization group, guided by our knowledge of four-dimensional non-Abelian gauge theories.
The running of the coupling $e^2(q)$ is governed by the beta function,
\begin{equation}
\frac{de^2}{d\log q}=N_fb_1 e^4/q+\cdots,
\label{RG1}
\end{equation}
where $b_1>0$ is the one-loop coefficient.
Upon defining a dimensionless coupling $g^2(q)=e^2/q$, we have
\begin{equation}
\frac{dg^2}{d\log q}=-g^2+N_fb_1 g^4+\cdots,
\label{RG2}
\end{equation}
a form typical of a super-renormalizable field theory.%
\footnote{$b_1$ depends on the renormalization scheme:
It is not invariant under a perturbative change in scheme, $g^2=g'^2+bg'^4+\cdots$.
Hence we do not specify it further, until we adopt the Schr\"odinger functional scheme below.
Since $b_1$ comes from screening by physical particles, it should be positive in any sensible  scheme.}
When $N_f$ is small, there is a region around $g^2=0$ where the running is dominated by the first term, which has the sign associated with asymptotic freedom in four dimensions.
As a consequence, the infrared running is towards strong coupling.
When the coupling becomes large enough, but before the second term becomes important, a condensate $\svev{\bar\psi\psi}$ forms.
This generates a mass for the fermions, whereupon at sufficiently small energy scales the fermions drop out of the theory, leaving free photons. The running of $e^2$ stops.

When $N_f$ is large, on the other hand, the one-loop term in \Eq{RG2} has to be considered.%
\footnote{In the 't~Hooft limit, $N_f\to\infty$ with $e^2N_f$ fixed, the one-loop term in \Eq{RG1} is the exact beta function.}
The beta function starts out negative at small $g^2$ but the one-loop term then makes it cross zero at $g^2=g_*^2\equiv(N_fb_1)^{-1}$.
This is an infrared-attractive fixed point.
Once the coupling runs there, a scale invariance sets in.
There are no massive particles and indeed, since Green functions behave as powers of distance, there are no particles at all.

There must be a value of $N_f$, which we have called $N_c$, that divides the two domains.
For $N_f$ just below this $N_c$, the one-loop term is significant:  It make the beta function turn towards zero as $g^2$ grows.
This only happens, however, at a strong coupling, and the condensate is triggered before $g^2$ actually reaches the zero of the one-loop beta function.
Then the fermions develop a mass and decouple in the infrared, as described for the small-$N_f$ case.

These infrared scenarios are familiar from non-Abelian theories in four dimensions.
In an asymptotically free theory the one-loop term in the beta function is negative, while the two-loop term changes sign as the number of fermion flavors is increased~\cite{Caswell:1974gg,Banks:1981nn}.
The transition from confinement to conformality has been much studied, most recently by the methods of lattice gauge theory (for surveys see~\cite{Neil:2012cb,Giedt:2012it}).

The Schr\"odinger functional (SF) method gives a non-perturbative definition of the beta function that lends itself to lattice calculations~\cite{Luscher:1992an,Luscher:1993gh}.
Originally designed for QCD \cite{Sint:1995ch,Jansen:1998mx,DellaMorte:2004bc}, it has also been applied to gauge theories that are on the borderline between confining and conformal \cite{Appelquist:2007hu,Shamir:2008pb,Appelquist:2009ty,Hietanen:2009az,Bursa:2009we,Bursa:2010xn,Hayakawa:2010yn,Karavirta:2011zg,DeGrand:2010na,DeGrand:2011qd,DeGrand:2012yq,DeGrand:2012qa,DeGrand:2013uha}.
In this paper we present a study of QED3 with $N_f=2$ four-component fermions by this method.
We first calculate and plot results for the beta function at fixed lattice spacing.
These indicate that the beta function deviates from the one-loop form (\ref{RG2}) at strong coupling and does not cross zero.
This is confirmed when we extrapolate to the continuum by two different methods.
Unless this trend is reversed at yet stronger couplings, our results imply that the $N_f=2$ theory is a confining theory with a mass scale.

As mentioned above, more straightforward lattice methods have also been applied to QED3.
We note in particular the extensive studies of Refs.~\cite{Hands:2002dv,Hands:2004bh,Strouthos:2008kc}.
This work has pointed towards $N_c\approx2$.
Calculating mass spectra and the chiral condensate is then quite challenging in the $N_f=2$ theory, because of the strong finite-volume effects inherent in low dimensionality: If $N_c$ is nearby then the condensate (if any) could be very small and particle masses (if any) likewise.
Deciding whether these quantities are really nonzero requires simulation on very large lattices~\cite{Gusynin:2003ww}.
The strength of the SF method, on the other hand, comes from the idea behind the renormalization group.
There is no need to control a large range of length scales in a given calculation, because the method relates nearby scales to derive the beta function.
Thus finite size turns from a hindrance to a basis for calculation.

This paper is organized as follows.
In Sec.~\ref{sec:SF} we review the SF method by adapting it to QED3;
the language is entirely that of the continuum.
We describe the lattice theory in Sec.~\ref{sec:lattice}: a non-compact Abelian gauge field coupled to $N_f$=2 four-component fermions.
We simulate this theory to obtain the running coupling, with results presented in Sec.~\ref{sec:beta}.
Section~\ref{sec:conclusions} contains further discussion of our method, while the appendices contain some technical details.

\section{Defining the running coupling\label{sec:SF}}
The Schr\"odinger functional offers a definition of the running coupling
that is convenient for non-perturbative lattice calculations.
We apply the method to QED3 in the continuum; lattice modifications will be given below.
The action is
\begin{equation}
S=\int d^3x\,\left(\frac1{4e_0^2}F_{\mu\nu}F^{\mu\nu}
+\sum_{i=1}^{N_f}\bar\psi_i \Sl D\psi_i\right),
\label{cont_action}
\end{equation}
where $F_{\mu\nu}=\partial_\mu A_\nu - \partial_\nu A_\mu$ is the usual field strength and $D_\mu=\partial_\mu+iA_\mu$.
With this definition, $A$ has dimensions of mass and so does the coupling $e_0^2$.
Renormalizing at a scale $\mu$, we identify $e_0^2\equiv e^2(\mu)$ and
define a dimensionless coupling $g^2(\mu)=e^2(\mu)/\mu$.

We work in a cubic volume of size $L^3$ and impose a background field in a way
that $L$ is its only scale.
Then calculation of the quantum effective action yields a coupling that runs with $L$, which we denote by $g^2(L)$.

We impose the background field through boundary conditions on the spacelike components $A_x$ and $A_y$ at
$t=0$ and $t=L$,
\begin{eqnarray}
  A_x &=& A_y \ = \ +\phi/L \,, \qquad t=0 \,,
\nonumber\\
  A_x &=& A_y \ = \ -\phi/L \,, \qquad t=L \,.
\label{bc}
\end{eqnarray}
The background field in the bulk is found by minimizing
the classical action,
\begin{equation}
S_{\text{cl}}=\frac1{e_0^2}\int d^3x\,\frac14 F_{\mu\nu}F^{\mu\nu},
\label{classical_action}
\end{equation}
and it is easy to see that the solution is a linear function of $t$,
\begin{equation}
A_x(t)=A_y(t)=\frac\phi L\left(1-\frac{2t}L\right),
\label{Afield}
\end{equation}
corresponding to a constant electric field,
\begin{equation}
E_x=E_y=-\frac{2\phi}{L^2}.
\label{Efield}
\end{equation}
The classical action of this field is
\begin{equation}
S_{\text{cl}}=\frac{4\phi^2}{e_0^2L}\equiv \frac{\tK(\phi)}{e_0^2L}.
\label{defK}
\end{equation}
In our work we fixed the background field parameter to be $\phi=\pi/4$.

The effective action $\Gamma=-\log Z$ gives a definition of the
running coupling in the SF scheme.
We write
\begin{equation}
\Gamma \equiv \frac1{e^2(L)}\int d^3x\,\frac14 F_{\mu\nu}F^{\mu\nu},
\end{equation}
where $F_{\mu\nu}$ is the classical background field.
Using Eqs.~(\ref{classical_action}) and~(\ref{defK}) gives
\begin{equation}
\Gamma=\frac{\tK(\phi)}{e^2(L)L}=\frac{\tK(\phi)}{g^2(L)}\,,
\end{equation}
and thus a calculation of $\Gamma$, in a three-dimensional volume of size $L$, gives directly  the running coupling $g^2(L)$ at the scale $L$.
In any field theory, one generally calculates not $\Gamma$ but its derivatives, which are given by Green functions.
Thus we differentiate with respect to the boundary parameter to obtain
\begin{equation}
\frac{\partial\Gamma}{\partial\phi}=\frac{K(\phi)}{g^2(L)}\,,
\label{GammaK}
\end{equation}
where $K = \partial\tK/\partial\phi$.
(Our choice $\phi=\pi/4$ gives $K=8\phi=2\pi$.)

We can check our numerical results by comparing to the first two terms in
a loop expansion for $\Gamma$,
\begin{equation}
\Gamma=S_{\text{cl}}+\Gamma^{(1)}+\cdots.
\end{equation}
The one-loop quantum correction is given by
\begin{equation}
\Gamma^{(1)}=-N_f\tr\log(\Sl D),
\end{equation}
and thus we define the one-loop quantity $c(\phi)$ via
\begin{equation}
\frac{\partial\Gamma^{(1)}}{\partial\phi}\equiv N_fK(\phi)c(\phi).
\label{NKc}
\end{equation}
[We have taken out a factor of $K(\phi)$ for convenience.]
Here $c(\phi)$ is a dimensionless function of the boundary conditions,
independent of the coupling and of the system size $L$.
We calculate it in Appendix \ref{app:oneloop}.
Inserting in \Eq{GammaK} we have the perturbative expression for the
running coupling,
\begin{equation}
\frac1{g^2(L)}=\frac1{g^2(\mu)\mu L}+N_fc+\cdots.
\label{RGsoln}
\end{equation}
Setting $q=1/L$, we rewrite the renormalization group equation (\ref{RG2}) as
\begin{equation}
\tbeta(1/g^2)\equiv\frac{d(1/g^2)}{d\log L}=-\frac1{g^2}+N_fb_1 +O(g^2).
\label{RG3}
\end{equation}
Upon differentiating \Eq{RGsoln}, we identify
\begin{equation}
b_1=c,
\end{equation}
the one-loop coefficient in the SF renormalization scheme.

\section{The lattice theory\label{sec:lattice}}

We use a non-compact formulation of the gauge field, wherein we define
the vector potential $A_{n\mu}$ on each link $(n,\mu)$ of the three-dimensional
lattice; we put a four-component
Dirac field $\psi_n$ on  each site $n$.
(We suppress the flavor index $f=1,2$ throughout.)
The Euclidean action contains the usual quadratic term for the gauge field and
a smoothed Wilson--clover action for the fermions,
\begin{equation}
S=\frac{\beta}2\sum_{\scriptstyle n\atop\scriptstyle \mu<\nu}(\nabla\times A)_{n\mu\nu}^2
+\bar\psi D\psi
\label{action}
\end{equation}
All the fields in \Eq{action} have been made dimensionless via appropriate
powers of the lattice spacing $a$.
The lattice curl is
\begin{equation}
(\nabla\times A)_{n\mu\nu}=A_{n\mu}+A_{n+\hat\mu,\nu}-A_{n+\hat\nu,\mu}-A_{n\nu},
\label{curl}
\end{equation}
and the first summation in \Eq{action} counts each plaquette once.
The fermion term is
\begin{subequations}
\begin{eqnarray}
\bar\psi D\psi&=&\sum_n\bar\psi_n\psi_n\label{action-1}\\
&&+\kappa\sum_{n\mu}\left[\bar\psi_n(1+\gamma_\mu)V_{n\mu}\psi_{n+\hat\mu}\right.\nonumber\\
&&\quad\left.+\bar\psi_{n+\hat\mu}(1-\gamma_\mu)V_{n\mu}^\dag\psi_n\right]\label{action-2}\\
&&+\kappa\csw\sum_n\bar\psi_n\frac i4\sigma_{\mu\nu}F_{n\mu\nu}\psi_n.\label{action-3}
\end{eqnarray}
\label{actionF}
\end{subequations}
The Wilson hopping term (\ref{action-2}) contains a link connection $V_{n\mu}$.
This is constructed from the compact gauge variables
\begin{equation}
U_{n\mu}=e^{iA_{n\mu}}
\end{equation}
by a normalized hypercubic (nHYP) smearing process~ \cite{Hasenfratz:2001hp,Hasenfratz:2007rf}, where each $V_{n\mu}$ is a weighted average of
the $U$ variables on nearby links (see Appendix \ref{app:smear}).
The purpose of this smearing is the suppression of lattice artifacts.
It allows us to go to stronger couplings before encountering numerical instabilities~\cite{DeGrand:2010na}.

A further cancellation of lattice artifacts is offered by the clover term (\ref{action-3})~\cite{Sheikholeslami:1985ij}.
While the field strength $F_{n\mu\nu}$ could be defined via the simple curl
(\ref{curl}), we adopt a definition appropriate to the compact theory,
a sum over the four leaves of the ``clover'' surrounding the site $n$,
\begin{equation}
F_{n\mu\nu}=\frac14\left(F^{(1)}_{n\mu\nu}+F^{(1)}_{n-\hat\mu,\mu\nu}+F^{(1)}_{n-\hat\nu,\mu\nu}+F^{(1)}_{n-\hat\mu-\hat\nu,\mu\nu}\right),
\end{equation}
where each term $F^{(1)}$ is a compact curl,
\begin{equation}
F^{(1)}_{n\mu\nu}=\sin(\nabla\times A)_{n\mu\nu}.
\end{equation}
This clover structure is the same as in non-Abelian theories and enables easy
adaptation of existing code.
We set the coefficient to its tree-level value, $\csw=1$, since we have found
this to be close to optimal when nHYP smearing is used~\cite{Shamir:2010cq}.

The boundary conditions (\ref{bc}) for the Schr\"odinger functional are now imposed on the gauge field on the time slices of the lattice at $t=0$ and $t=L$.
Moreover, there are no dynamical fermion fields on these boundaries.

The coupling in \Eq{action} is
\begin{equation}
\beta=\frac1{e_0^2a},
\end{equation}
where $e_0$ is the bare charge.
The hopping parameter is related to the bare electron mass $m_0$ by
\begin{equation}
\kappa=\frac1{6+2m_0a}.
\end{equation}
We study the massless theory by demanding that the measured, physical
fermion mass $m$ vanish.
We define this from the axial Ward identity,
\begin{equation}
\partial_\mu^- A_\mu^a=2m P^a,
\label{AWI}
\end{equation}
where $A^a_\mu=\frac12\bar\psi\gamma_\mu\gamma_5\tau^a\psi$ is the isovector axial current and $P^a=\frac12\bar\psi\gamma_5\tau^a\psi$ is the isovector
pseudoscalar density.  ($\partial^-_\mu$ is the backward
lattice derivative.)
It is convenient in a SF calculation to define an gauge-invariant, pseudoscalar
wall source operator ${\cal O}^a$ near the boundary at $t=0$ (see Refs.~\cite{Shamir:2010cq,Luscher:1996sc} for details).
Then \Eq{AWI} can be used to relate two Green functions at zero spatial
momentum, viz.,
\begin{equation}
\partial_0^-\sum_{\bf x}\svev{A^a_0({\bf x},t){\cal O}^a}
=2m\sum_{\bf x}\svev{P^a({\bf x},t){\cal O}^a}.
\end{equation}
We evaluate the Green functions at $t=L/2$, whence the ratio gives $m$.
At each value of $\beta$, we tune the hopping parameter $\kappa$ to make $m$
vanish, which defines the critical curve $\kappa_c(\beta)$.
We list the values of $\beta$ used in this study, as well as the values of $\kappa_c(\beta)$, in Table~\ref{tab:Kc}.

As mentioned above, a Monte Carlo simulation of the lattice theory does not give the effective action directly.
Instead, one applies the Schr\"odinger functional method by using Green functions, which are derivatives of $\Gamma$.
The derivative on the left-hand side of Eq.~(\ref{GammaK}) is
\begin{equation}
  \left.\frac{\partial \Gamma}{\partial\phi} \right|_{\phi=\pi/4}
  =
  \left.\svev{\frac{\partial S_{G}}{\partial\phi}
  -\tr \left( \frac{1}{D^\dagger}\;
        \frac{\partial (D^\dagger D)}{\partial\phi}\;
            \frac{1}{D} \right)}\right|_{\phi=\pi/4},
\label{dphi}
\end{equation}
where $S_G$ is the pure gauge action, the first term in \Eq{action}.
Equation (\ref{dphi}) is a particular expectation value of the gauge fields and the Dirac operator $D$.
Differentiating with respect to $\phi$
is the same as differentiating
with respect to $A_{n\mu}$ on the boundary.
We also impose twisted spatial boundary conditions on the fermion fields~\cite{Sint:1995ch},
$\psi(x+L)=\exp(i\theta)\psi(x)$, with $\theta=\pi/5$ on both spatial axes.

We employed the hybrid Monte Carlo (HMC) algorithm~\cite{Duane:1985hz,Duane:1986iw,Gottlieb:1987mq} in our simulations.  The molecular dynamics integration was accelerated with an
additional heavy pseudo-fermion field~\cite{Hasenbusch:2001ne}, multiple time
scales~\cite{Urbach:2005ji}, and a second-order Omelyan
integrator~\cite{Takaishi:2005tz}.
Since the non-compact formulation allows gauge fluctuations in which $A_{n\mu}$ can wander to infinity, we monitor the field and carry out gauge transformations, local and global, to keep the field within certain bounds.
\begin{table}
\begin{ruledtabular}
\begin{tabular}{lcrrrr}
 & & \multicolumn{4}{c}{trajectories (thousands)}\\
\cline{3-6}
$\beta$ & $\kappa_{c}$ & $L=8a$ & $L=12a$ & $L=16a$ & $L=24a$\\
\hline
0.355 & 0.17440 & 150 & 150 & --  & --\\
0.39  & 0.17325 & 150 & 100 & --  & --\\
0.4   & 0.17282 & 100 & 150 & 150 & 122\\
0.458 & 0.17166 & 100 & 100 & --  & --\\
0.541 & 0.17055 & 100 & 100 & --  & --\\
0.6   & 0.16994 & 100 & 100 & 150 & 110\\
0.8   & 0.16898 & 100 & 100 & 100 & 100\\
1     & 0.16848 & 100 & 100 & 100 & 105\\
2     & 0.16757 & --  & --  & 50  & 50
\end{tabular}
\end{ruledtabular}
\caption{Summary of simulation runs for obtaining the running coupling $g^{2}$
at bare couplings $\left(\beta,\kappa_{c}\right)$ for lattice sizes
$L$ used in this work. We used a trajectory length of unity for most
of the simulations. The exceptions are the strong coupling runs ($\beta\le0.4$)
and the $\beta=0.6$ run on the biggest lattice, in which we used a
trajectory length of $1/2$.
\label{tab:Kc} }
\end{table}
\begin{table}
\begin{ruledtabular}
\begin{tabular}{lcccc}
 & \multicolumn{4}{c}{$1/g^{2}$}\\
\cline{2-5}
$\beta$ &  $L=8a$ & $L=12a$ & $L=16a$ & $L=24a$\\
\hline
0.355 & 0.0753(20) & 0.0515(35) & --         & --\\
0.39  & 0.0751(18) & 0.0608(42) & --         & --\\
0.4   & 0.0753(24) & 0.0591(34) & 0.0558(41) & 0.0422(75)\\
0.458 & 0.0824(22) & 0.0644(41) & --         & --\\
0.541 & 0.0992(16) & 0.0771(21) & --         & --\\
0.6   & 0.1110(16) & 0.0833(22) & 0.0686(21) & 0.0485(63)\\
0.8   & 0.1374(17) & 0.1018(22) & 0.0854(26) & 0.0635(32)\\
1     & 0.1643(16) & 0.1258(21) & 0.0978(25) & 0.0837(32)\\
2     & --         & --         & 0.1732(35) & 0.1277(43)
\end{tabular}
\end{ruledtabular}
\caption{Schr\"odinger functional running coupling calculated at bare couplings $\beta$ on lattices of size $L$.  The first two lines are affected by lattice artifacts that make them unusable in calculating the RDBF.
\label{tab:schrodinger-func}}
\end{table}

\section{The running coupling and the beta function \label{sec:beta}}

We calculated the running coupling from Eqs.~(\ref{GammaK}) and~(\ref{dphi}) for lattice sizes
$L/a=8,$ 12, 16, and 24 for the bare couplings listed in Table~\ref{tab:Kc}.
The results are shown in Table \ref{tab:schrodinger-func}.
We use different subsets of the data for two different analysis methods.
Our goal is the beta function $\tbeta$, defined in \Eq{RG3}, or, equivalently, its representation as the rescaled discrete beta function (RDBF) \cite{DeGrand:2011qd} for scaling by a fixed factor $s$,
\begin{equation}
R(u,s)\equiv\frac{u(sL)-u(L)}{\log s},
\label{RDBF}
\end{equation}
where the argument is
\begin{equation}
u=u(L)\equiv\frac{1}{g^{2}\left(L\right)}.
\end{equation}
It is clear that at a fixed point, where $\tbeta=0$, the RDBF will be zero as well.
\subsection{Discrete beta function---two-lattice matching}
\begin{figure}
\begin{center}
\includegraphics[width=\columnwidth,clip]{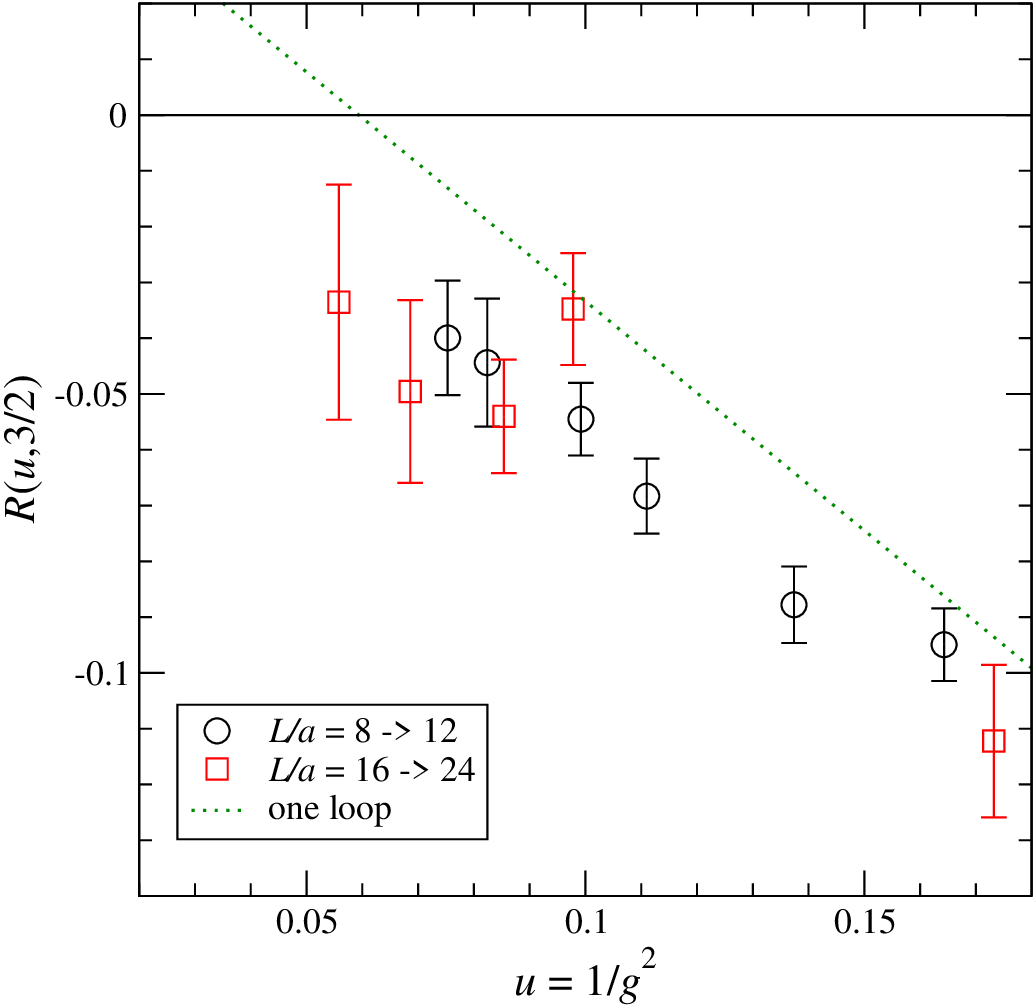}
\end{center}
\caption{The rescaled discrete beta function calculated from each pair of couplings for the
lattices of size $L/a=8\rightarrow12$ and $16\rightarrow24$.
\label{fig:RDBF}}
\end{figure}

We calculate the RDBF directly by comparing pairs of lattice sizes $L$ and $L'=sL$ at fixed bare coupling $\beta$.
[Two lattices with the same $(\beta,\kappa)$ have the same lattice spacing.]
Lattices of size $L=8a$ and~$12a$ give a scale factor $s=3/2$, as do lattices of size $L=16a$ and $24a$.
We plot the RDBF for all such pairs of lattices in Fig.~\ref{fig:RDBF}.
For comparison we plot the one-loop formula, derived from using \Eq{RGsoln} in the definition (\ref{RDBF}),
\begin{equation}
R^{(1)}(u,s)=\frac{1-1/s}{\log s}\left(-u+N_fc\right).
\end{equation}
We note the general trend that as $u$ decreases the data deviate downwards from the one-loop curve---away from the axis---and avoid its zero.

\begin{figure}
\begin{center}
\includegraphics[width=\columnwidth,clip]{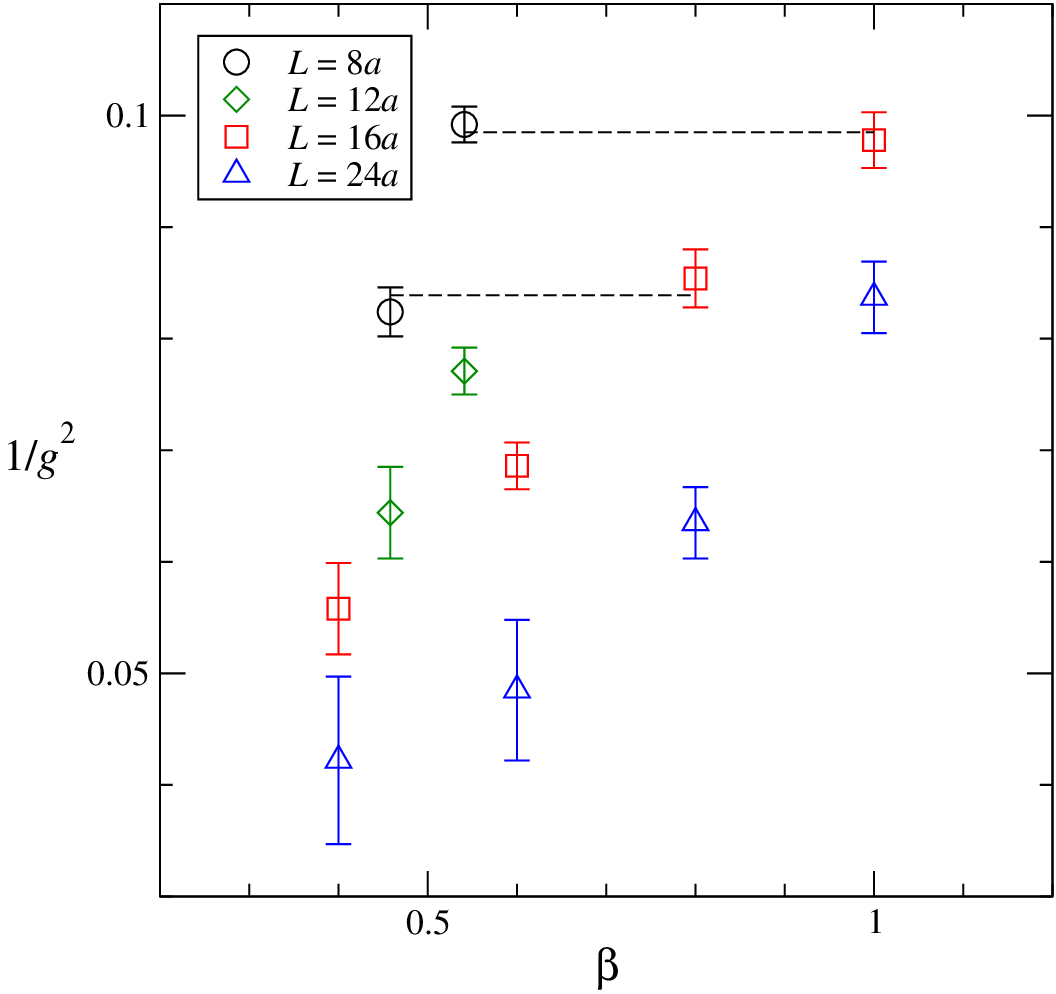}
\end{center}
\caption{Extrapolating the RDBF to the continuum.
Plotted is a subset of the results for the running coupling that are listed in Table~\ref{tab:schrodinger-func}.
Pairs of data points at the same $\beta$ give the RDBF for a single lattice spacing.
The horizontal lines link data points at the same coupling $u=1/g^2$ but at different lattice spacings.
These pairs give the RDBF at the two lattice spacings, to be compared and extrapolated in Fig.~\ref{fig:RDBF2}.
\label{fig:two-lattice}}
\end{figure}

Figure~\ref{fig:RDBF} is a first look only, since the dependence on lattice spacing has not yet been studied.
The quantity plotted in Fig.~\ref{fig:RDBF} is really $R(u,s;a)$, where the added argument is the lattice spacing.
To extrapolate to $a=0$, we seek data for $R$ at fixed $u$---which means fixed $L$---but at different $a$, and thus different lattice size $L/a$.
We show this procedure in Fig.~\ref{fig:two-lattice}.
\begin{figure}
\begin{center}
\includegraphics[width=\columnwidth,clip]{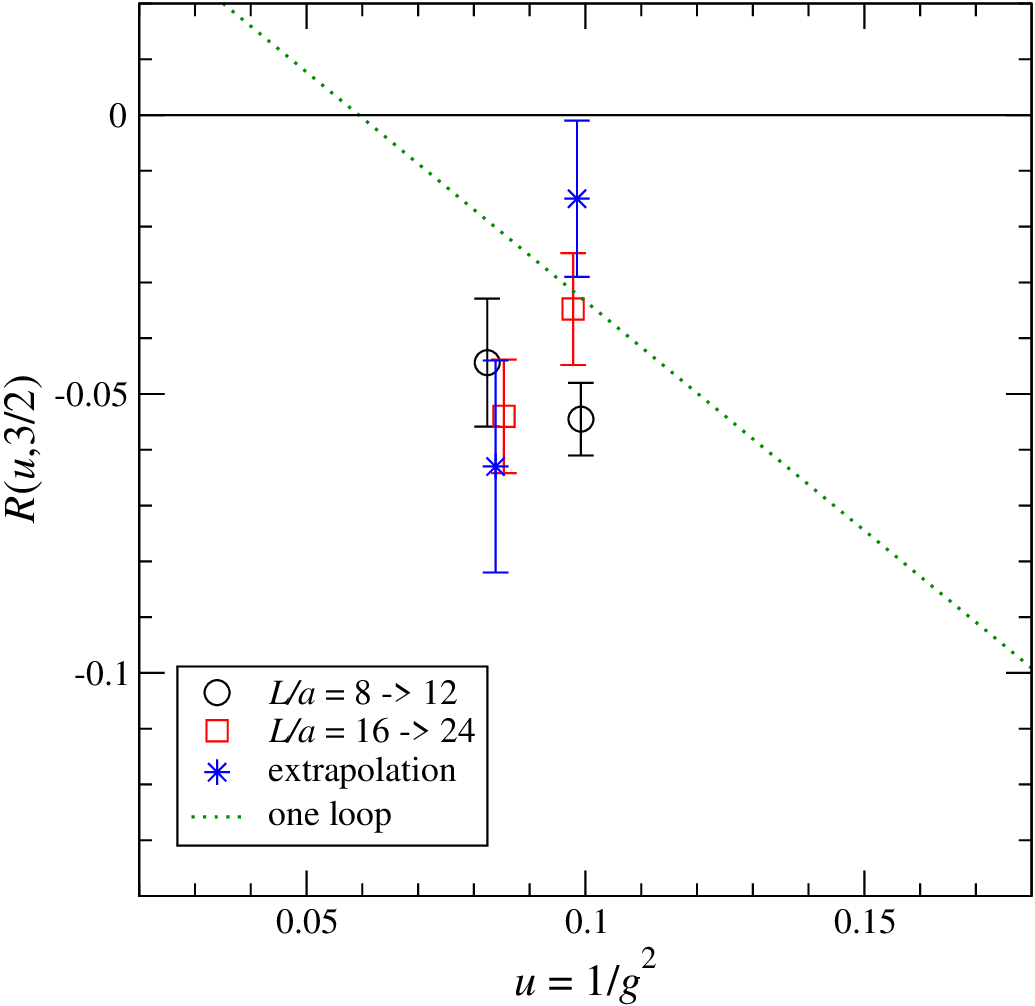}
\end{center}
\caption{Extrapolation of the RDBF to the continuum, at the two couplings for which we have the needed data.
\label{fig:RDBF2}}
\end{figure}
The horizontal lines link data points at fixed $u$, but measured on different lattice sizes $L/a=8$ and~16.%
\footnote{The points for $L=8a$ were calculated at $\beta=0.458$ and~$0.541$---lines~4
and~5 of Table~\ref{tab:schrodinger-func}.
These $\beta$ values were chosen for this purpose, i.e., to match $u$ to the $L=16a$ values at $\beta=0.8$ and~1.}
Thus the RDBFs calculated from these points are $R(u,s;a=L/8)$ and $R(u,s;a=L/16)$, respectively.
We have such pairs at fixed $u$ at $u\simeq0.84$ and $u\simeq0.98$.
We replot them in Fig.~\ref{fig:RDBF2}, together with the extrapolations according to
\begin{equation}
R(u,s;a)=R(u,s;a=0)+C\frac aL.
\end{equation}

It proved impossible to extend the two-lattice matching method to stronger coupling because of lattice artifacts.
We attempted to extend the $L=8a$ data to stronger couplings than those shown in Fig.~\ref{fig:two-lattice} (see Table~\ref{tab:schrodinger-func}).
The measured coupling $1/g^2$ levels off, rather than continuing downward with smaller $\beta$.
This prevents matching to the $L=16a$ data.

\begin{figure}
\begin{center}
\includegraphics[width=\columnwidth,clip]{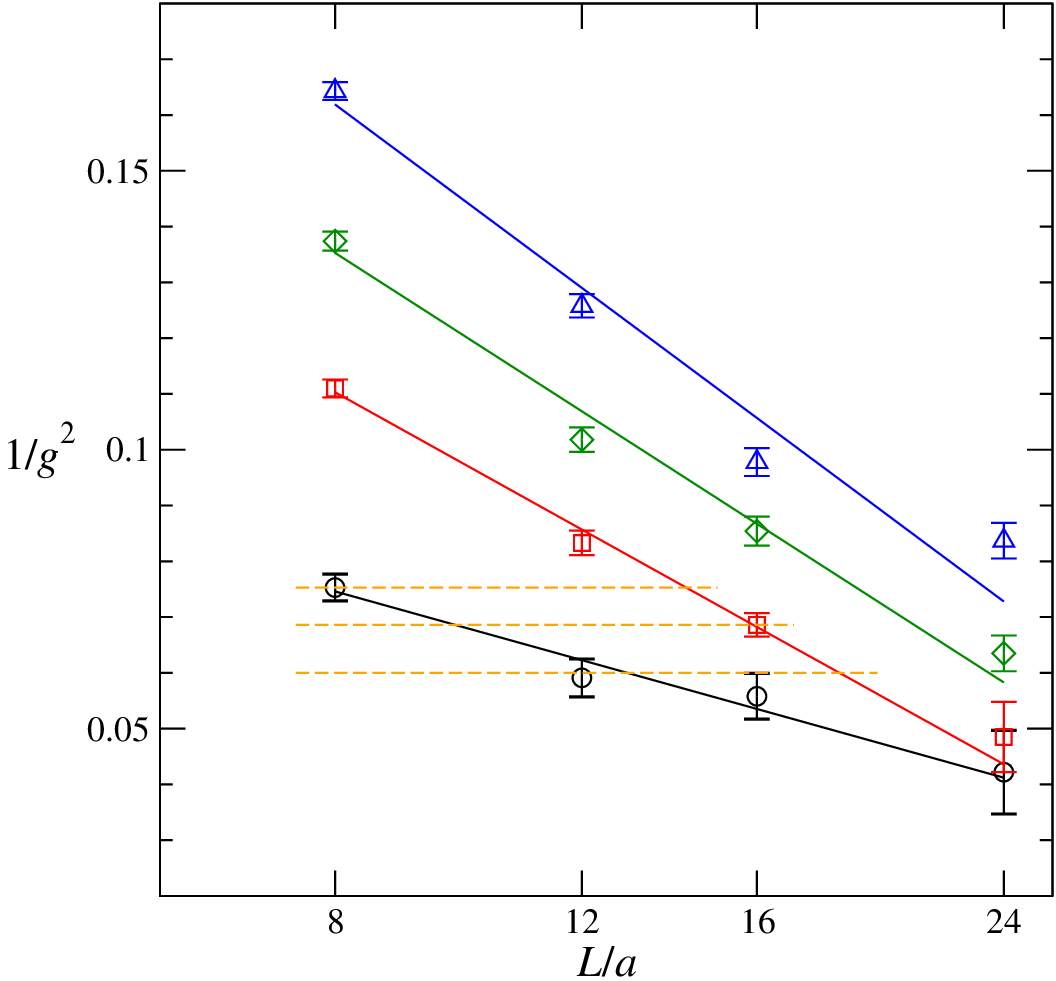}
\end{center}
\caption{Running couplings $1/g^2$ vs $\log(L/a)$, from Table~\ref{tab:schrodinger-func}.
Each set corresponds to a fixed bare coupling $\beta$.
Top to bottom: $\beta=1$, 0.8, 0.6,~0.4.
The solid lines are the linear fits (\ref{l1fit}) for
each bare coupling.
The horizontal, dashed lines are values of $1/g^2$ chosen for extrapolating the slopes to $a/L=0$.
\label{fig:sfB1}}
\end{figure}

\subsection{Slope analysis}
In order to perform an analysis at stronger couplings than the two-lattice method allows, we begin by plotting data sets for $1/g^2$ in Fig.~\ref{fig:sfB1}.
At four values of $\beta$, we have results for $1/g^2$ on four lattice volumes, $L=8a$, $12a$, $16a$, and~$24a$.
Since fixing $\beta$ fixes the lattice spacing $a$, these sets represent scaling in $L$ at fixed $a$.
As $L$ is increased, $1/g^2$ decreases in accord with a negative beta function.
The figure shows the results of a straight-line fit to each data set,
\begin{equation}
u(L)=\frac{1}{g^{2}(L)}=c_{0}+c_{1}\log L/a.
\label{l1fit}
\end{equation}
At the two weakest bare couplings, $\beta=1.0$ and~0.8, the fit is poor because of the obvious curvature of the dependence on $\log (L/a)$.
This means that the decrease in $1/g^2$ is not described by a constant beta function $\tbeta$.
Indeed, the one-loop continuum formula predicts that the decrease in $1/g^2$ slows as one moves leftward in Fig.~\ref{fig:RDBF}.

At $\beta=0.6$ and~0.4, on the other hand, a straight line fits the points well, with perhaps a small deviation for $L=24a$ where the error bar is large anyway.
Since $g^2$ is {\em not\/} constant as $L$ changes, this motivates the hypothesis that the beta function $\tbeta(u)$ has leveled off in the range of $u$ covered by these data.
Thus we arrive at estimates $\tbeta(u)\simeq c_1$.
The different slopes of the two data sets show that the estimates vary with the lattice spacing.

\begin{figure}
\begin{center}
\includegraphics[width=\columnwidth,clip]{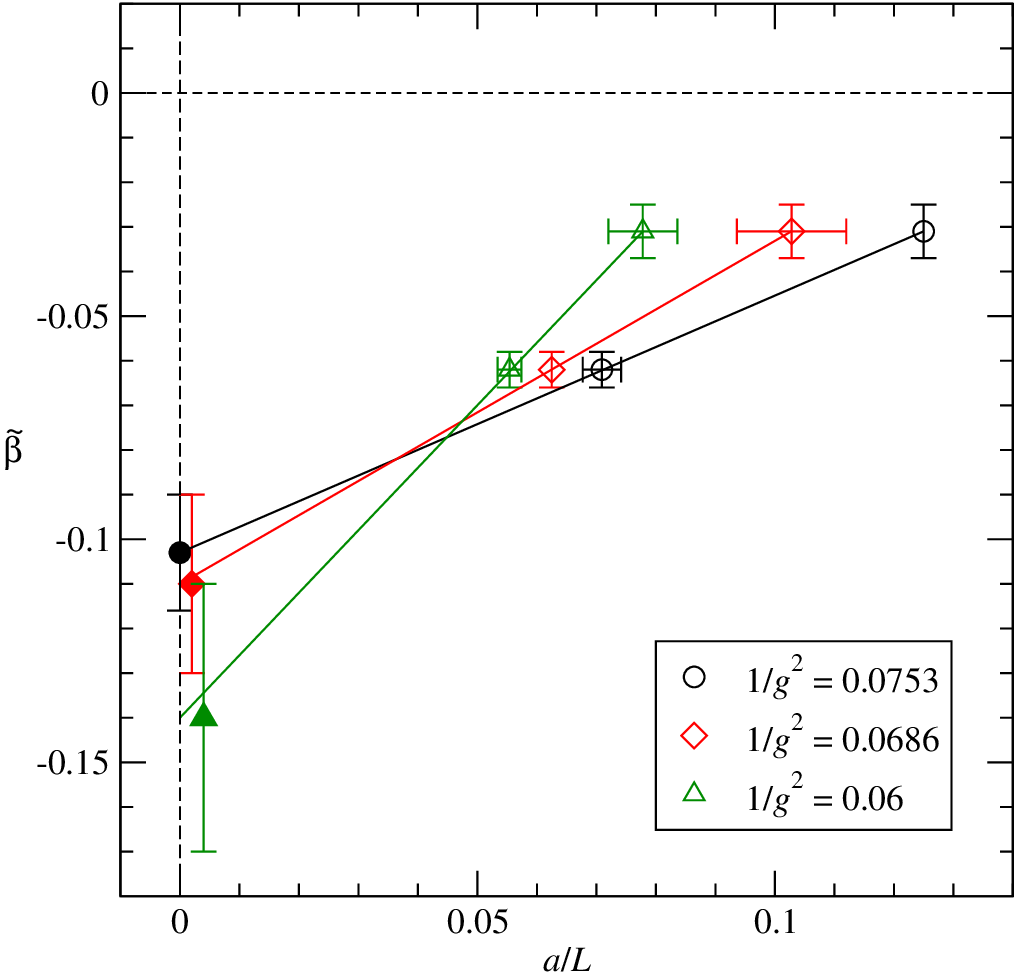}
\end{center}
\caption{Extrapolating the beta function to the continuum, as described in Table~\ref{tab:slopes}.
The $y$-values of the upper and lower sets of data are the slopes of the fits shown in Fig.~\ref{fig:sfB1} for $\beta=0.4$ and $0.6$.
The $x$-values are the inverses of the intercepts of the horizontal lines in that figure with the fit lines.
The data are extrapolated to $a/L=0$ as shown (solid points, displaced horizontally from the axis for clarity).
\label{fig:tbeta_ex}}
\end{figure}

\begin{figure}
\begin{center}
\includegraphics[width=\columnwidth,clip]{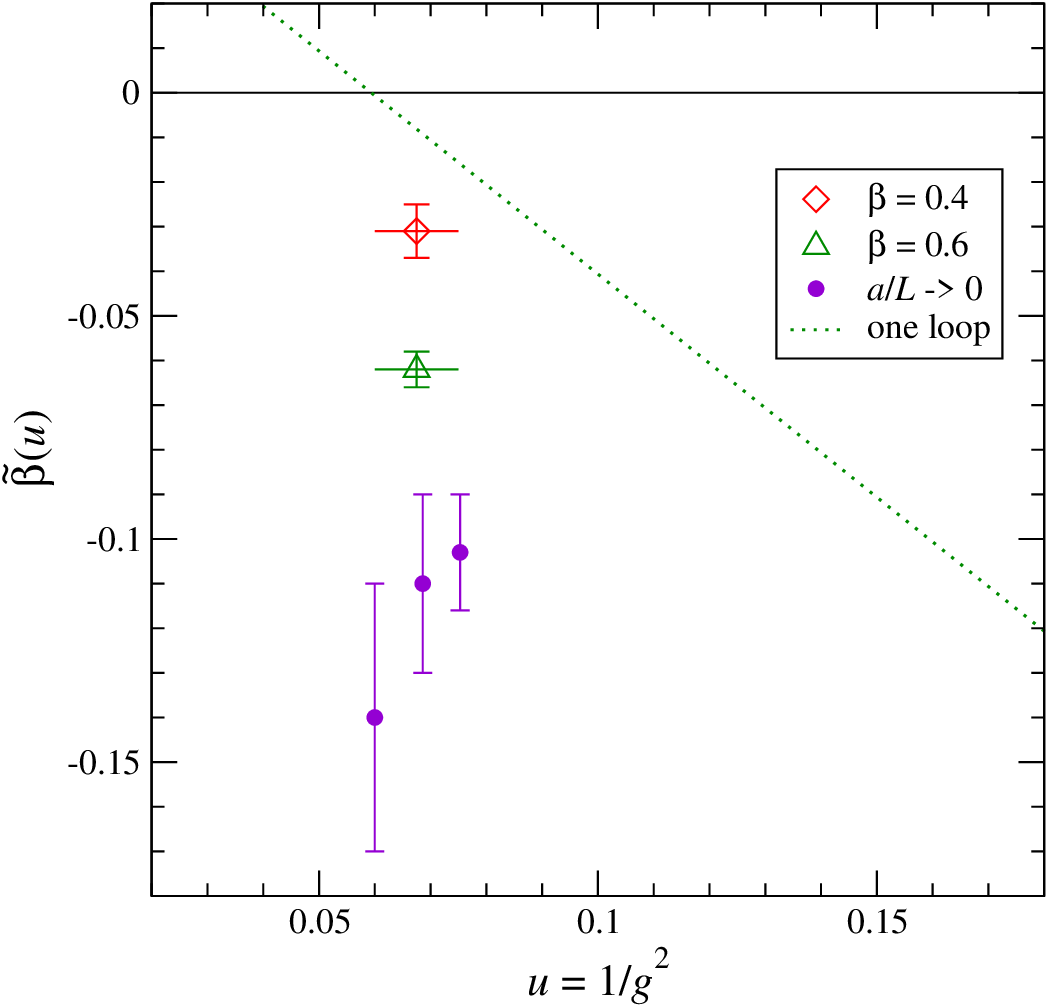}
\end{center}
\caption{Extrapolation of the beta function to the continuum (Table~\ref{tab:slopes} and Fig.~\ref{fig:tbeta_ex}), plotted against the coupling.
The points labeled $\beta=0.4$ and~0.6 are the $y$-values of the data plotted in Fig.~\ref{fig:tbeta_ex}.
Assuming the slopes in Fig.~\ref{fig:sfB1} to be constant in the coupling range shown gives a set of extrapolations to $a/L=0$, depending on the chosen value of $u$, as shown.
The extrapolations are not statistically independent of each other.
\label{fig:tbeta}}
\end{figure}

Choosing a value of $u=1/g^2$ in Fig.~\ref{fig:sfB1} gives a horizontal line that intersects the fit lines for the two data sets.
The two intercepts give two values of $a/L$, one for each fit.
We again extrapolate these to the continuum linearly,
\begin{equation}
\tbeta(u;a)=\tbeta(u;a=0)+C\frac aL,
\end{equation}
as shown in Table~\ref{tab:slopes} and Fig.~\ref{fig:tbeta_ex}.
The final result is plotted in Fig.~\ref{fig:tbeta}.

The three $u$-values chosen span a short interval.
We do not go to yet stronger coupling in this analysis because lowering the chosen value of $u$ in Fig.~\ref{fig:sfB1} will give intercepts that are too far to the right; here the data deviate from the straight lines, as they must, and the straight lines will not give accurate values of $\tbeta$.

[A complementary argument is the following.  The quality of the linear fits in Fig.~\ref{fig:sfB1}
is good for both $\beta=0.4$ and 0.6.
Therefore, when we extrapolate to the continuum limit
it is legitimate to use the two slopes as estimates for the beta functions
for any value $0.0422 \le u \le 0.0753$,
which is the interval covered by the $\beta=0.4$ data.
(It is contained in the interval covered by the $\beta=0.6$ data.)
As we move down in $u$, however, the intercepts with the linear fits
get closer to each other, and, moreover, the uncertainty in the $L/a$ value of each intercept grows.  As a result, the uncertainty in
the continuum extrapolation grows rapidly, a trend that is clear in Figs.~\ref{fig:tbeta_ex} and~\ref{fig:tbeta}.
Thus no further information will emerge from pushing to stronger coupling.]

Figure~\ref{fig:tbeta} shows that our numerical results deviate considerably from the one-loop curve.
They show no sign of crossing the axis;
taking Figs.~\ref{fig:RDBF2} and~\ref{fig:tbeta} together, we have a beta function that approaches the axis with the one-loop curve but then curves away from it in the strong coupling region.
\begin{table}
\begin{ruledtabular}
\begin{tabular}{llll}
$1/g^2$     &  $L/a(\beta=0.4)$ & $L/a(\beta=0.6)$ & $\tbeta(a/L\to0)$\\
\hline
0.0753(24)  & \ 8              & 14.10(64)      & $-0.103(13)$     \\
0.0686(21)  & \ 9.73(87)       & 16             & $-0.11(2)$       \\
0.06        & 12.85(96)        & 18.04(65)      & $-0.14(3)$
\end{tabular}
\end{ruledtabular}
\caption{Extrapolation to $a/L=0$ in slope analysis.
$1/g^2$ is the chosen value of the running coupling for the horizontal line in Fig.~\ref{fig:sfB1}.
The next two columns are the values of $L/a$ for the intercepts of the horizontal line with the fit lines for $\beta=0.4$ and~0.6.
The slopes for the two lines, which give $\tbeta(u;a)$, are respectively -0.031(6) and -0.062(4).
The final column gives the extrapolation from these two values to $a/L=0$ (Figs.~\ref{fig:tbeta_ex} and~\ref{fig:tbeta}).
\label{tab:slopes}}
\end{table}
\section{Discussion \label{sec:conclusions}}
QED3 is qualitatively different from the four-dimensional non-Abelian theories studied with the SF method~\cite{Appelquist:2007hu,Shamir:2008pb,Appelquist:2009ty,Hietanen:2009az,Bursa:2009we,Bursa:2010xn,Hayakawa:2010yn,Karavirta:2011zg,DeGrand:2010na,DeGrand:2011qd,DeGrand:2012yq,DeGrand:2012qa,DeGrand:2013uha}.
This is evident already in the weak-coupling behavior of the beta function.
In four dimensions, the beta function $\tbeta(u)$ for the inverse coupling is constant in one loop; in three, it is linear in $u$, as seen in \Eq{RG3} and in the figures.
Moreover, in the borderline-conformal theories there is a partial cancellation between the one-loop and two-loop terms that makes the beta function small compared to that of QCD.
These differences are reflected in the analysis technique we adopt here, which is different from that in our earlier papers
\cite{DeGrand:2011qd,DeGrand:2012yq,DeGrand:2012qa,DeGrand:2013uha}
and more closely resembles the applications of the method to QCD~\cite{DellaMorte:2004bc}.

Our technique of linear fits at fixed bare coupling $\beta$ [\Eq{l1fit} and Fig.~\ref{fig:sfB1}] was designed for a slowly running theory, where the slopes are small and the coupling changes little from the smallest lattice to the largest.
It works best when the beta function changes slowly as well.
It is plain that neither property holds in Fig.~\ref{fig:sfB1}:
the slopes are large and the couplings change rapidly.
For the two largest values of $\beta$, where the beta function is largest, the rate of change of the coupling decreases markedly as one follows the data to large volumes, so that the non-constancy of the beta function is evident.
At the smaller $\beta$'s the slopes hold more nearly constant, validating the fits and allowing us to  make the hypothesis that the beta function has levelled off.

We have used two different methods for extrapolation to the continuum, each with its own limitations.
In weak coupling, we used the two-lattice method to extrapolate the rescaled discrete beta function.
This works as long as one can match couplings between two different lattice sizes.
As we saw, it fails at stronger couplings because of lattice artifacts.
At the stronger bare couplings we use the fact that, as in QCD, the data in Fig.~\ref{fig:sfB1} cover overlapping ranges in the running coupling.
The overlap between $\beta=0.6$ and~0.4 allows for a straightforward extrapolation of the slope---the beta function---to $a/L=0$.

It is evident from  Fig.~\ref{fig:sfB1} that the slopes decrease as $\beta$ decreases, as is to be expected from the form of the one-loop beta function.
Thus the horizontal lines in Fig.~\ref{fig:sfB1} will associate the larger lattices (in terms of $L/a$) with larger slopes.
It is then inevitable that extrapolating the slopes in Fig.~\ref{fig:tbeta_ex} pushes the beta function away from the axis.
Is the result, then, due more to our method than to the data?
The examples offered by the four-dimensional non-Abelian theories show that this is not the case~\cite{DeGrand:2011qd,DeGrand:2012yq,DeGrand:2012qa,DeGrand:2013uha}.
There, as one approaches a possible zero of the beta function, the slopes become so small that the fixed-$\beta$ data sets do not overlap in coupling.
This prevents an extrapolation of slopes to the continuum limit in those theories.
An alternative method, based only the linear fits at fixed $\beta$, leads to a beta function that stays near (or even crosses) zero~\cite{DeGrand:2012yq}.

\begin{acknowledgments}
We thank N.~Itzhaki for urging us to work on QED3.
B.S. thanks the UCLA Physics Department for its hospitality when this work was begun.
This work was supported in part by the Israel Science Foundation under Grants~423/09
and~1362/08 and by the European Research Council under Grant~203247.
Our lattice simulation program had its origins in the publicly available package of the MILC collaboration
\cite{MILC} with nHYP smearing adapted from code by A.~Hasenfratz, R.~Hoffmann, and S.~Schaefer~\cite{Hasenfratz:2001tw}.
\end{acknowledgments}
\appendix
\section{The running coupling in one loop \label{app:oneloop}}

As explained in Sec.~\ref{sec:SF}, the one-loop coefficient in the beta function is given by
$b_1=c$, where
\begin{equation}
c=-\frac{1}{K}\frac{\partial}{\partial\phi}\log\det\Sl D\,,
\label{KcD}
\end{equation}
which is to be evaluated at $\phi=\pi/4$, our choice for the SF boundary
conditions.

The calculation of the derivative of the fermion determinant
using the lattice regularization follows
closely Ref.~\cite{Sint:1995ch}.
Since $\Dslash$ is defined in a uniform (but time-dependent) background potential (\ref{Afield}),
its eigenfunctions take the form
\begin{equation}
\psi_{\alpha}(n)=\exp\left(i {\bf p} \cdot {\bf x}\right)f_{\alpha}(t),
\end{equation}
where ${\bf p}=(p_1,p_2)$, and $\alpha$ is a Dirac index.
The allowed spatial momenta are
\begin{equation}
p_{k}=\left(2\pi n_{k}+\theta\right)/L, \qquad k=1,2,
\end{equation}
where $\theta$ is the fermion twist angle, and $n_{k}=1,\ldots,\ell$,
with $\ell=L/a$.  Thus the calculation is reduced to
\begin{equation}
c = -\frac{1}{K} \lim_{\ell\to\infty} \sum_{\bf p}
\frac{\partial}{\partial\phi}\log\det\tilde{\Sl D}({\bf p}),
\label{sumpk}
\end{equation}
where $\tilde{\Sl D}({\bf p})$ is a $4(\ell-1)\times4(\ell-1)$ matrix
acting on functions of time $f_\alpha(t)$.  (We recall that in the SF setup,
the dynamical fermionic degrees of freedom live on
time slices $t/a=1,\ldots, \ell-1$.)
As shown in Ref.~\cite{Sint:1995ch}, one can simplify this expression to
\begin{equation}
\frac{\partial}{\partial\phi}\log\det\tilde{\Dslash}=\Tr\left(M^{-1}\frac{\partial}{\partial\phi}M\right),
\label{DM}
\end{equation}
where $M=M({\bf p})$ is a $2\times2$ matrix that encodes the hopping
in the time direction.  For a fixed lattice of size $\ell$, the calculation
proceeds by using Eq.~(\ref{DM}) and summing over all ${\bf p}$.
Thanks to the clover term, the result rapidly tends to a constant
as $L$ is increased, and we find
\begin{equation}
c=0.0297(1) \ .
\end{equation}

\begin{figure}
\begin{center}
\includegraphics[width=\columnwidth,clip]{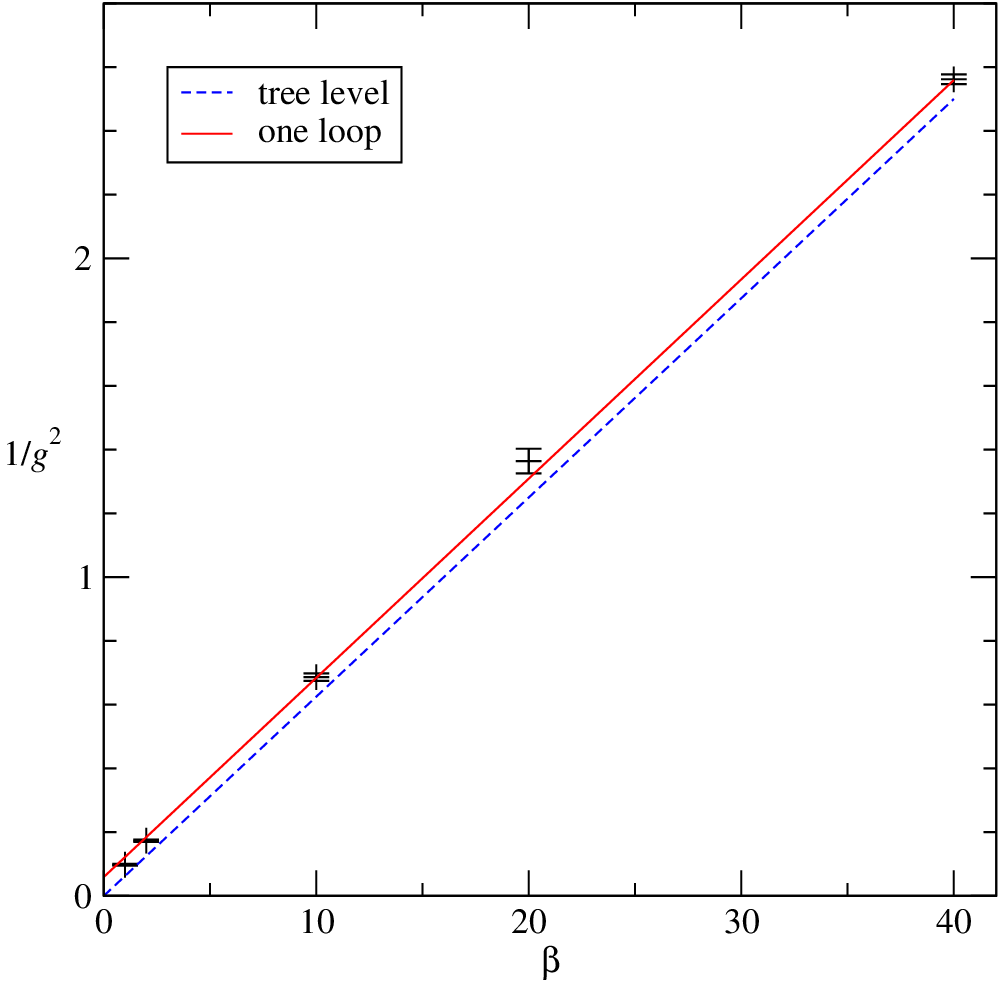}
\end{center}
\caption{Simulation results for $L=16a$ in the weak coupling regime
compared to the calculated tree-level and one-loop expressions.
\label{fig:weak}}
\end{figure}

A comparison of the one-loop calculation with lattice simulations
at very weak bare coupling is shown in Fig.~\ref{fig:weak}.
There is good agreement between the simulation results and
the one-loop expression
\begin{equation}
\frac{1}{g^{2}(L)}=\frac{\beta}{\ell}+N_{f}c \ .
\end{equation}
\section{$\textbf{n}$HYP smearing for the Abelian theory\label{app:smear}}
We modify the formulas of nHYP smearing
\cite{Hasenfratz:2001hp,Hasenfratz:2007rf,DeGrand:2012qa} for the U(1)
gauge group.
While the smearing could be simplified by using the non-compact gauge
field $A_{n\mu}$, we can take advantage of our experience with non-Abelian
theories (as well as of existing code) by modifying the non-Abelian
formulas to use the compact field $U_{n\mu}$.
The reduction  in dimensionality to three shortens the smearing procedure.
The smeared links $V_{n\mu}$ are constructed from the bare links $U_{n\mu}$
in two consecutive smearing steps via intermediate fields $\widetilde{V}$
according to
\begin{widetext}
\begin{subequations}
\begin{eqnarray}
V_{n\mu} & = & \textrm{Norm}\left[(1-\alpha_{1})U_{n\mu}
+\frac{\alpha_{1}}{4}\sum_{\pm\nu\neq\mu}\widetilde{V}_{n\nu;\mu}\widetilde{V}_{n+\hat{\nu},\mu;\nu}\widetilde{V}_{n+\hat{\mu},\nu;\mu}^{\dagger}\right],\label{eq:HYP-def1}\\
\widetilde{V}_{n\mu;\nu} & = & \textrm{Norm}\left[(1-\alpha_{2})U_{n\mu}
+\frac{\alpha_{2}}{2}\sum_{\pm\eta\neq\nu,\mu}U_{n\eta}U_{n+\hat{\eta},\mu}U_{n+\hat{\mu},\eta}^{\dagger}\right].
\end{eqnarray}
\label{eq:HYP-def}
\end{subequations}
\end{widetext}
Each step is a weighted averaging of a link with staples surrounding it.
The restricted sums ensure that only links that share a cube with
$U_{n\mu}$ enter the smearing.
We set the parameters $\alpha_{1,2} =(0.75,0.6)$ as used in Refs.~\cite{Hasenfratz:2001hp,Hasenfratz:2007rf} for the four-dimensional theory (where there is
an additional direction in which to smear).
The normalizations
indicated in Eqs.~(\ref{eq:HYP-def}) are simply
\begin{equation}
\textrm{Norm}\,V=\frac{V}{|V|}.
\end{equation}



\begin{thebibliography}{99}


\bibitem{Pisarski:1984dj}
  R.~D.~Pisarski,
  ``Chiral Symmetry Breaking in Three-Dimensional Electrodynamics,''
  Phys.\ Rev.\ D {\bf 29}, 2423 (1984).

\bibitem{Appelquist:2004ib}
  T.~Appelquist and L.~C.~R.~Wijewardhana,
  ``Phase structure of noncompact QED3 and the Abelian Higgs model,''
  hep-ph/0403250.

\bibitem{Appelquist:1986fd} 
  T.~W.~Appelquist, M.~J.~Bowick, D.~Karabali and L.~C.~R.~Wijewardhana,
  ``Spontaneous Chiral Symmetry Breaking in Three-Dimensional QED,''
  Phys.\ Rev.\ D {\bf 33}, 3704 (1986).
  
\bibitem{Appelquist:1988sr} 
  T.~Appelquist, D.~Nash and L.~C.~R.~Wijewardhana,
  ``Critical Behavior in (2+1)-Dimensional QED,''
  Phys.\ Rev.\ Lett.\  {\bf 60}, 2575 (1988).

\bibitem{Nash:1989xx} 
  D.~Nash,
  ``Higher Order Corrections in (2+1)-Dimensional QED,''
  Phys.\ Rev.\ Lett.\  {\bf 62}, 3024 (1989).

\bibitem{Pennington:1990bx} 
  M.~R.~Pennington and D.~Walsh,
  ``Masses from nothing: A Nonperturbative study of QED in three-dimensions,''
  Phys.\ Lett.\ B {\bf 253}, 246 (1991).
  
\bibitem{Curtis:1992gm} 
  D.~C.~Curtis, M.~R.~Pennington and D.~Walsh,
  ``Dynamical mass generation in QED in three-dimensions and the $1/N$ expansion,''
  Phys.\ Lett.\ B {\bf 295}, 313 (1992).

\bibitem{Appelquist:1994ui} 
  T.~Appelquist, J.~Terning and L.~C.~R.~Wijewardhana,
  ``(2+1)-dimensional QED and a novel phase transition,''
  Phys.\ Rev.\ Lett.\  {\bf 75}, 2081 (1995)
  [hep-ph/9402320].

\bibitem{Ebihara:1994wm} 
  T.~Ebihara, T.~Iizuka, K.~-i.~Kondo and E.~Tanaka,
  ``Dynamical breakdown of chirality and parity in (2+1)-dimensional QED,''
  Nucl.\ Phys.\ B {\bf 434}, 85 (1995)
  [hep-ph/9404361].

\bibitem{Maris:1996zg} 
  P.~Maris,
  ``The Influence of the full vertex and vacuum polarization on the fermion propagator in QED in three-dimensions,''
  Phys.\ Rev.\ D {\bf 54}, 4049 (1996)
  [hep-ph/9606214].

\bibitem{Aitchison:1997ua} 
  I.~J.~R.~Aitchison, N.~E.~Mavromatos and D.~McNeill,
  ``Inverse Landau--Khalatnikov transformation and infrared critical exponents of (2+1)-dimensional quantum electrodynamics,''
  Phys.\ Lett.\ B {\bf 402}, 154 (1997)
  [hep-th/9701087].

\bibitem{Gusynin:1998kz} 
  V.~P.~Gusynin, V.~A.~Miransky and A.~V.~Shpagin,
  ``Effective action and conformal phase transition in three-dimensional QED,''
  Phys.\ Rev.\ D {\bf 58}, 085023 (1998)
  [hep-th/9802136].

\bibitem{Appelquist:1999hr}
  T.~Appelquist, A.~G.~Cohen and M.~Schmaltz,
  ``A new constraint on strongly coupled gauge theories,''
  Phys.\ Rev.\ D {\bf 60}, 045003 (1999)
  [hep-th/9901109].

\bibitem{Gusynin:2003ww} 
  V.~P.~Gusynin and M.~Reenders,
  ``Infrared cutoff dependence of the critical flavor number in three-dimensional QED,''
  Phys.\ Rev.\ D {\bf 68}, 025017 (2003)
  [hep-ph/0304302].
  
\bibitem{Fischer:2004nq} 
  C.~S.~Fischer, R.~Alkofer, T.~Dahm and P.~Maris,
  ``Dynamical chiral symmetry breaking in unquenched QED$_3$,''
  Phys.\ Rev.\ D {\bf 70}, 073007 (2004)
  [hep-ph/0407104].

\bibitem{Kaveh:2004qa} 
  K.~Kaveh and I.~F.~Herbut,
  ``Chiral symmetry breaking in three-dimensional quantum electrodynamics in presence of irrelevant interactions: A Renormalization group study,''
  Phys.\ Rev.\ B {\bf 71}, 184519 (2005)
  [cond-mat/0411594].

\bibitem{Bashir:2008fk} 
  A.~Bashir, A.~Raya, I.~C.~Cloet and C.~D.~Roberts,
  ``Regarding confinement and dynamical chiral symmetry breaking in QED3,''
  Phys.\ Rev.\ C {\bf 78}, 055201 (2008)
  [arXiv:0806.3305 [hep-ph]].

\bibitem{Goecke:2008zh} 
  T.~Goecke, C.~S.~Fischer and R.~Williams,
  ``Finite volume effects and dynamical chiral symmetry breaking in QED3,''
  Phys.\ Rev.\ B {\bf 79}, 064513 (2009)
  [arXiv:0811.1887 [hep-ph]].

\bibitem{Braun:2014wja} 
  J.~Braun, H.~Gies, L.~Janssen and D.~Roscher,
  ``On the Phase Structure of Many-Flavor QED$_3$,''
  arXiv:1404.1362 [hep-ph].

\bibitem{Dagotto:1988id} 
  E.~Dagotto, J.~B.~Kogut and A.~Koci\'c,
  ``A Computer Simulation of Chiral Symmetry Breaking in (2+1)-Dimensional QED with $N$ Flavors,''
  Phys.\ Rev.\ Lett.\  {\bf 62}, 1083 (1989).

\bibitem{Dagotto:1989td} 
  E.~Dagotto, A.~Koci\'c and J.~B.~Kogut,
  ``Chiral Symmetry Breaking in Three-dimensional {QED} With $N_f$ Flavors,''
  Nucl.\ Phys.\ B {\bf 334}, 279 (1990).

\bibitem{Azcoiti:1993ey} 
  V.~Azcoiti and X.~-Q.~Luo,
  ``Phase structure of compact lattice QED in three-dimensions with massless Fermions,''
  Mod.\ Phys.\ Lett.\ A {\bf 8}, 3635 (1993)
  [hep-lat/9212011].
  
\bibitem{Alexandre:2001pa} 
  J.~Alexandre, K.~Farakos, S.~J.~Hands, G.~Koutsoumbas and S.~E.~Morrison,
  ``Three-dimensional QED with dynamical fermions in an external magnetic field,''
  Phys.\ Rev.\ D {\bf 64}, 034502 (2001)
  [hep-lat/0101011].

\bibitem{Lee:2002id} 
  D.~Lee and P.~Maris,
  ``Massless three-dimensional QED with explicit fermions,''
  Phys.\ Rev.\ D {\bf 67}, 076002 (2003)
  [hep-lat/0212033].

\bibitem{Fiore:2005ps} 
  R.~Fiore, P.~Giudice, D.~Giuliano, D.~Marmottini, A.~Papa and P.~Sodano,
  ``QED(3) on a space-time lattice: Compact versus noncompact formulation,''
  Phys.\ Rev.\ D {\bf 72}, 094508 (2005)
  [Erratum-ibid.\ D {\bf 72}, 119902 (2005)]
  [hep-lat/0506020].

\bibitem{Hands:2002dv}
  S.~J.~Hands, J.~B.~Kogut and C.~G.~Strouthos,
  ``Non-compact QED$_3$ with $N_f\ge2$,''
  Nucl.\ Phys.\ B {\bf 645}, 321 (2002)
  [hep-lat/0208030].

\bibitem{Hands:2004bh}
  S.~J.~Hands, J.~B.~Kogut, L.~Scorzato and C.~G.~Strouthos,
  ``Non-compact three-dimensional quantum electrodynamics with $N_f = 1$ and $N_f = 4$,''
  Phys.\ Rev.\ B {\bf 70}, 104501 (2004)
  [hep-lat/0404013].

\bibitem{Strouthos:2008kc}
  C.~Strouthos and J.~B.~Kogut,
  ``The Phases of Non-Compact QED$_3$,''
  PoS LAT {\bf 2007}, 278 (2007)
  [arXiv:0804.0300 [hep-lat]].

\bibitem{Caswell:1974gg}
  W.~E.~Caswell,
  ``Asymptotic behavior of nonabelian gauge theories to two loop order,''
  Phys.\ Rev.\ Lett.\  {\bf 33}, 244 (1974).

\bibitem{Banks:1981nn}
  T.~Banks and A.~Zaks,
  ``On the phase structure of vector-like gauge theories with massless fermions,''
  Nucl.\ Phys.\  B {\bf 196}, 189 (1982).

\bibitem{Neil:2012cb}
  E.~T.~Neil,
  ``Exploring Models for New Physics on the Lattice,''
  PoS LATTICE {\bf 2011}, 009 (2011)
  [arXiv:1205.4706 [hep-lat]].

\bibitem{Giedt:2012it}
  J.~Giedt,
  ``Lattice gauge theory and physics beyond the standard model,''
  PoS LATTICE {\bf 2012}, 006 (2012).


\bibitem{Luscher:1992an}
  M.~L\"uscher, R.~Narayanan, P.~Weisz and U.~Wolff,
  ``The Schrodinger functional: A Renormalizable probe for nonAbelian gauge
  theories,''
  Nucl.\ Phys.\  B {\bf 384}, 168 (1992)
  [arXiv:hep-lat/9207009].

\bibitem{Luscher:1993gh}
  M.~L\"uscher, R.~Sommer, P.~Weisz and U.~Wolff,
  ``A precise determination of the running coupling in the SU(3) Yang-Mills theory,''
  Nucl.\ Phys.\  B {\bf 413}, 481 (1994)
  [arXiv:hep-lat/9309005].

\bibitem{Sint:1995ch}
  S.~Sint and R.~Sommer,
  ``The running coupling from the QCD Schr\"odinger functional: A one loop analysis,''
  Nucl.\ Phys.\  B {\bf 465}, 71 (1996)
  [arXiv:hep-lat/9508012].

\bibitem{Jansen:1998mx}
  K.~Jansen and R.~Sommer  [ALPHA collaboration],
  ``O($\alpha$) improvement of lattice QCD with two flavors of Wilson quarks,''
  Nucl.\ Phys.\  B {\bf 530}, 185 (1998)
  [Erratum-{\em ibid.}\  B {\bf 643}, 517 (2002)]
  [arXiv:hep-lat/9803017].

\bibitem{DellaMorte:2004bc}
  M.~Della Morte {\em et al.} [ALPHA Collaboration],
  ``Computation of the strong coupling in QCD with two dynamical flavours,''
  Nucl.\ Phys.\  B {\bf 713}, 378 (2005)
  [arXiv:hep-lat/0411025].

\bibitem{Appelquist:2007hu}
  T.~Appelquist, G.~T.~Fleming and E.~T.~Neil,
  \ttl{Lattice study of the conformal window in QCD-like theories,}
  Phys.\ Rev.\ Lett.\  {\bf 100}, 171607 (2008)
  [Erratum-ibid.\  {\bf 102}, 149902 (2009)]
  [arXiv:0712.0609 [hep-ph]].

\bibitem{Appelquist:2009ty}
  T.~Appelquist, G.~T.~Fleming and E.~T.~Neil,
  \ttl{Lattice Study of Conformal Behavior in SU(3) Yang-Mills Theories,}
  Phys.\ Rev.\ D {\bf 79}, 076010 (2009)
  [arXiv:0901.3766 [hep-ph]].

\bibitem{Hietanen:2009az}
  A.~J.~Hietanen, K.~Rummukainen and K.~Tuominen,
  \ttl{Evolution of the coupling constant in SU(2) lattice gauge theory with two adjoint fermions,}
  Phys.\ Rev.\ D {\bf 80}, 094504 (2009)
  [arXiv:0904.0864 [hep-lat]].

\bibitem{Bursa:2009we}
  F.~Bursa, L.~Del Debbio, L.~Keegan, C.~Pica and T.~Pickup,
  \ttl{Mass anomalous dimension in SU(2) with two adjoint fermions,}
  Phys.\ Rev.\  D {\bf 81}, 014505 (2010)
  [arXiv:0910.4535 [hep-ph]].

\bibitem{Bursa:2010xn}
  F.~Bursa, L.~Del Debbio, L.~Keegan, C.~Pica and T.~Pickup,
  \ttl{Mass anomalous dimension in SU(2) with six fundamental fermions,}
  Phys.\ Lett.\ B {\bf 696}, 374 (2011)
  [arXiv:1007.3067 [hep-ph]].

\bibitem{Hayakawa:2010yn}
  M.~Hayakawa, K.~-I.~Ishikawa, Y.~Osaki, S.~Takeda, S.~Uno and N.~Yamada,
  \ttl{Running coupling constant of ten-flavor QCD with the Schr\"odinger functional method,}
  Phys.\ Rev.\ D {\bf 83}, 074509 (2011)
  [arXiv:1011.2577 [hep-lat]].

\bibitem{Karavirta:2011zg}
  T.~Karavirta, J.~Rantaharju, K.~Rummukainen and K.~Tuominen,
  \ttl{Determining the conformal window: SU(2) gauge theory with $N_f = 4$, 6 and 10 fermion flavours,}
  arXiv:1111.4104 [hep-lat];
%
  ``Exploring the conformal window: SU(2) gauge theory on the lattice,''
  arXiv:1201.2037 [hep-lat].

\bibitem{Shamir:2008pb}
  Y.~Shamir, B.~Svetitsky and T.~DeGrand,
  ``Zero of the discrete beta function in SU(3) lattice gauge theory with color sextet fermions,''
  Phys.\ Rev.\  D {\bf 78}, 031502 (2008)
  [arXiv:0803.1707 [hep-lat]].

\bibitem{DeGrand:2010na}
  T.~DeGrand, Y.~Shamir, B.~Svetitsky,
  ``Running coupling and mass anomalous dimension of SU(3) gauge theory with two flavors of symmetric-representation fermions,''
  Phys.\ Rev.\  {\bf D82}, 054503 (2010)
  [arXiv:1006.0707 [hep-lat]].

\bibitem{DeGrand:2011qd}
  T.~DeGrand, Y.~Shamir, B.~Svetitsky,
  ``Infrared fixed point in SU(2) gauge theory with adjoint fermions,''
  Phys.\ Rev.\  {\bf D83}, 074507 (2011)
  [arXiv:1102.2843 [hep-lat]].

\bibitem{DeGrand:2012yq}
  T.~DeGrand, Y.~Shamir and B.~Svetitsky,
  ``Mass anomalous dimension in sextet QCD,''
  arXiv:1201.0935 [hep-lat].

\bibitem{DeGrand:2012qa}
  T.~DeGrand, Y.~Shamir and B.~Svetitsky,
  ``SU(4) lattice gauge theory with decuplet fermions: Schrodinger functional analysis,''
  Phys.\ Rev.\ D {\bf 85}, 074506 (2012)
  [arXiv:1202.2675 [hep-lat]].

\bibitem{DeGrand:2013uha} 
  T.~DeGrand, Y.~Shamir and B.~Svetitsky,
  ``Near the sill of the conformal window: gauge theories with fermions in two-index representations,''
  Phys.\ Rev.\ D {\bf 88}, no. 5, 054505 (2013)
  [arXiv:1307.2425].


\bibitem{Hasenfratz:2001hp}
  A.~Hasenfratz and F.~Knechtli,
  ``Flavor symmetry and the static potential with hypercubic blocking,''
  Phys.\ Rev.\  D {\bf 64}, 034504 (2001)
  [arXiv:hep-lat/0103029].

\bibitem{Hasenfratz:2007rf}
  A.~Hasenfratz, R.~Hoffmann and S.~Schaefer,
  ``Hypercubic smeared links for dynamical fermions,''
  JHEP {\bf 0705}, 029 (2007)
  [arXiv:hep-lat/0702028].

\bibitem{Sheikholeslami:1985ij}
  B.~Sheikholeslami and R.~Wohlert,
  ``Improved continuum limit lattice action for QCD with Wilson fermions,''
  Nucl.\ Phys.\  B {\bf 259}, 572 (1985).

\bibitem{Luscher:1996sc} 
  M.~Luscher, S.~Sint, R.~Sommer and P.~Weisz,
  ``Chiral symmetry and $O(a)$ improvement in lattice QCD,''
  Nucl.\ Phys.\ B {\bf 478}, 365 (1996)
  [hep-lat/9605038].

\bibitem{Shamir:2010cq}
  Y.~Shamir, B.~Svetitsky and E.~Yurkovsky,
  ``Improvement via hypercubic smearing in triplet and sextet QCD,''
  Phys.\ Rev.\ D {\bf 83}, 097502 (2011)
  [arXiv:1012.2819 [hep-lat]].

\bibitem{Duane:1985hz}
  S.~Duane and J.~B.~Kogut,
  ``Hybrid Stochastic Differential Equations Applied to Quantum Chromodynamics,''
  Phys.\ Rev.\ Lett.\  {\bf 55}, 2774 (1985).

\bibitem{Duane:1986iw}
  S.~Duane and J.~B.~Kogut,
  ``The Theory Of Hybrid Stochastic Algorithms,''
  Nucl.\ Phys.\ B {\bf 275}, 398 (1986).

\bibitem{Gottlieb:1987mq}
  S.~A.~Gottlieb, W.~Liu, D.~Toussaint, R.~L.~Renken and R.~L.~Sugar,
  ``Hybrid Molecular Dynamics Algorithms for the Numerical Simulation of Quantum Chromodynamics,''
  Phys.\ Rev.\ D {\bf 35}, 2531 (1987).

\bibitem{Hasenbusch:2001ne}
  M.~Hasenbusch,
  ``Speeding up the Hybrid-Monte-Carlo algorithm for dynamical fermions,''
  Phys.\ Lett.\  B {\bf 519}, 177 (2001)
  [arXiv:hep-lat/0107019].

\bibitem{Urbach:2005ji}
 C.~Urbach, K.~Jansen, A.~Shindler and U.~Wenger,
 ``HMC algorithm with multiple time scale integration and mass preconditioning,''
 Comput.\ Phys.\ Commun.\  {\bf 174}, 87 (2006)
 [arXiv:hep-lat/0506011].

\bibitem{Takaishi:2005tz}
  T.~Takaishi and P.~de Forcrand,
  ``Testing and tuning new symplectic integrators for hybrid Monte Carlo algorithm in lattice QCD,''
  Phys.\ Rev.\  E {\bf 73}, 036706 (2006)
  [arXiv:hep-lat/0505020].

\bibitem{MILC} {\tt http://www.physics.utah.edu/$\sim$detar/milc/}

\bibitem{Hasenfratz:2001tw}
  A.~Hasenfratz, R.~Hoffmann and F.~Knechtli,
  ``The static potential with hypercubic blocking,''
  Nucl.\ Phys.\ Proc.\ Suppl.\  {\bf 106}, 418 (2002)
  [hep-lat/0110168].


\end{thebibliography}
\end{document}